\documentclass[showkeys,showpacs,amssymb,
eqsecnum,preprintnumbers,nofootinbib]{revtex4}
\usepackage{graphics}
\usepackage{latexsym}
\usepackage{amsfonts}
\usepackage{hyperref}
\usepackage{bm}

\begin{document}

\newcommand{\g}{g_{\mu\nu}}
\newcommand{\kk}{q}
\newcommand{\vp}{\varphi}
\newcommand{\remark}{{\bf ***}}
\newcommand{\mm}{{\bf ********}}
\newcommand{\nn}{\nonumber\\}
\newcommand{\hsi}{\hat{\sigma}}
\newcommand{\hrho}{\hat{\rho}}
\newcommand{\hl}{\hat{\lambda}_\k}
\newcommand{\si}{\sigma}
\newcommand{\mubar}{\bar{\mu}}
\newcommand{\dlt}{\bigtriangleup}
\newcommand{\beq}{\begin{equation}}
\newcommand{\eeq}{\end{equation}}
\newcommand{\bed}{\begin{displaymath}}
\newcommand{\eed}{\end{displaymath}}
\def\bea{\begin{eqnarray}}
\def\eea{\end{eqnarray}}


\title{Quantum fluctuations for de Sitter branes in bulk AdS$_5$}
\thanks {W.N. would like to dedicate this work to A. Naylor and S.
 Katsumoto for much love and support.}
\author{W. Naylor}
\email[Email: ]{naylor@yukawa.kyoto-u.ac.jp}
\affiliation{ Yukawa Institute for Theoretical Physics, Kyoto University,
Kyoto 606-8502, Japan}
\author{M. Sasaki}
\email[Email: ]{misao@yukawa.kyoto-u.ac.jp}
\affiliation{ Yukawa Institute for Theoretical Physics, Kyoto University,
Kyoto 606-8502, Japan}

\begin{abstract} 
 The vacuum expectation value of the square of the field fluctuations
 of a scalar field on a background consisting of {\it two} de Sitter
 branes embedded in an anti-de Sitter bulk are considered. 
We apply a dimensional reduction to obtain an effective lower dimensional
 de Sitter space equation of motion with associated Kaluza-Klein masses and 
canonical commutation relations.
The case of a scalar field obeying a restricted class of mass 
and curvature couplings, including massless, conformal coupling as a special
 case, is considered.
We find that the local behaviour of the quantum fluctuations suffers from
 surface divergences as we approach the brane, however, if the field is 
{\it constrained} to its value on the brane from the beginning then surface
 divergences disappear. 
The ratio of $\langle\phi^2\rangle$ between the Kaluza-Klein spectrum and 
the lowest eigenvalue mode
 is found to vanish in the limit that one of the branes goes to
 infinity.
\end{abstract}

\preprint{YITP-04-65}
\maketitle
\begin{center}
\today
\end{center}

\section{Introduction}

In this article we discuss the evaluation of the 
vacuum expectation value of the square of the field fluctuations 
(quantum fluctuations) on the background
which consists of positive and negative tension de Sitter branes 
embedded in an AdS$_5$ bulk. 
We shall use the following notation for quantum fluctuations;
 $\langle\phi^2(z)\rangle$,
where $z$ is the coordinate for the extra dimension and it
signifies that the quantity is a local functional of
 the extra dimension. 

Quantum fluctuations are of primary interest in inflationary cosmology,
 because they are related to the power spectrum, and thus, may have an
 observable effect on the cosmic microwave background.
 In the work of \cite{KKS,SHS} quantum 
fluctuations of a bulk scalar field
were investigated by canonical methods for massless and massive
 fields respectively. 
It was argued that in the limit $H\ell\ll 1$ (where $H$ is Hubble's constant
and $\ell$ the AdS$_5$ radius) it is physically reasonable to 
use $H$ as a natural cut off, even for the Kaluza-Klein modes. 
However, in the opposite limit the Kaluza-Klein contribution becomes
 significant. The work of \cite{KKS,SHS} looked at the difference between
 the Kaluza-Klein contributions and the bound state mode for field
 fluctuations constrained to the brane.
 However, because of the logarithmic divergence associated with the
integral over the Kaluza-Klein tower, there was an explicit cutoff
dependence in their result. Furthermore, they did not evaluate 
$\langle\phi^2(z)\rangle$ in the bulk. 
Thus, a particular motivation for this current work is to explicitly
 investigate the local effects of the bulk on this contribution
 (see Ref.~\cite{PT} for field perturbations in a slightly different
 brane world model). 

In this article we shall use the canonical formalism to obtain 
a dimensionally reduced theory in terms 
of a de Sitter wave equation of motion with associated Kaluza-Klein masses. 
Then, by choosing the Euclidean vacuum state we are able to evaluate
 the quantum fluctuations in terms 
of zeta functions. For simplicity we focus on a restricted class of mass 
and curvature couplings of the scalar field, because in this case
 the eigenfunction solutions are more manageable with eigenvalues that 
are explicitly known. 
Although we are mainly concerned with a 5-dimensional set up, much of
 the paper is given in $(D+1)$-dimensions. 

The layout of the paper is the following: 
In Sec.~\ref{canonics} we perform the dimensional reduction to 
an effective $D$-dimensional de Sitter space equation of motion.
In Sec.~\ref{zetphi} we discuss the relationship between the local zeta
 function and Green function.
In Sec.~\ref{spec} we considers a special case suitable for simple evaluation
of the local mode sums directly and go on to evaluate the quantum fluctuations
 for such a case in Sec.~\ref{flucts}. 
We close the article in Sec.~\ref{conc} with some discussions and defer 
the calculation of the lowest eigenvalue mode contribution and
the flat brane limit to Appendices~\ref{bstate} and \ref{flatapp},
 respectively.

\section{Dimensional Reduction}
\label{canonics}

In this section we start by performing the dimensional reduction of
a bulk scalar field in the background consisting of an AdS$_{D+1}$ 
bulk with two $D$-dimensional de Sitter branes embedded. 
For a clear account of the dimensional reduction procedure for
 flat bounding branes in an AdS bulk, see Ref.~\cite{FT},
 which also discusses the effects of induced curvature terms.

We consider a bulk scalar field with the action
\beq
S=\frac 1 2 \int d^{D+1}X
\sqrt{-g^{(D+1)}}~\phi(X)
\left(\Box_{D+1}-M^2-\xi R^{(D+1)}\right)\phi(X)\,,
\label{act}
\eeq
where the capital $X$ is used to denote the $(D+1)$-dimensional 
coordinates.
The $(D+1)$-dimensional bulk metric is given by
\beq
ds^2 =dr^2+b^2(r)ds_{\rm dS}^2\,; \qquad b(r)=H\ell\sinh(r/\ell),
\label{lormetr}
\eeq
where $ds_{\rm dS}^2$ is the $D$-dimensional de Sitter metric
with radius $H^{-1}$. Note that the choice of $H$ is arbitrary here.
The $(D+1)$-dimensional scalar curvature is given by
\beq
R^{(D+1)} =\frac{R^{(D)}}{b^2(r)}-\Delta(r),
\label{scal}
\eeq
with 
\beq
\Delta(r) = \left[2D {b''(r)\over b(r)}+
D(D-1) \left({b'(r)\over b(r)}\right)^2\right],
\eeq
and $R^{(D)}=D(D-1)H^2$ is the de Sitter scalar curvature.
The positive and negative tension branes are located at $r_\pm$,
 respectively.
The brane tensions are~\cite{GS,GEN}
\begin{equation}
\sigma_\pm=\pm\frac{2(D-1)}{\kappa^2\ell}\coth {r_\pm\over \ell}\,.
\label{tension}
\end{equation}

In the Heisenberg picture, the field operator $\phi(X)$
can be expressed by the Fock representation. Namely,
we solve the field equation to obtain a complete set 
of mode functions $u_\Lambda(X)$
(in the sense of the Klein-Gordon inner product),
expand $\phi$ in terms of $u_\Lambda$,
\begin{eqnarray}
\phi(X)=\sum_\Lambda \left(a_\Lambda u_\Lambda(X)
+a^*_\Lambda u^*_\Lambda(X)\right),
\label{modedecomp}
\end{eqnarray}
and quantise the coefficients $a_\Lambda$ and $a^*_\Lambda$,
where the complex conjugate becomes the Hermite conjugate operator,
 by imposing the canonical
commutation relation,
\begin{eqnarray}
\left[{a_\Lambda,a^\dag_{\Lambda'}}\right]=\delta_{\Lambda,\Lambda'}\,.
\end{eqnarray}
Then the vacuum expectation value $\langle\phi^2(X)\rangle$ is
formally given by
\begin{eqnarray}
\langle\phi^2(X)\rangle=\sum_\Lambda |u_\Lambda(X)|^2.
\label{phi2}
\end{eqnarray}

The method we employ here is based on a partial application of
this fact. Let us first briefly describe our method.
If we decompose the field $\phi$ in terms of a
 complete set of mode functions as given by Eq.~(\ref{modedecomp}),
we find it can be expressed in the form,
\begin{eqnarray}
\phi(X)=\sum_n u_{n}(r)\phi_{n}(x),
\label{drphi}
\end{eqnarray}
where $x$ denotes the $D$-dimensional coordinates,
$\phi_n(x)$ is a $D$-dimensional scalar field
satisfying the field equation,
\beq
\left[-\Box_D + (m_n^2+\rho^2)H^2\right]\phi_{n}(x^\mu)= 0,
\label{sphereq}
\eeq
and the radial solution $u_n(r)$ satisfies 
the eigenvalue equation
\bea
\left[{d^2\over dr^2}
+\frac{D}{\ell}\coth\left(\frac{r}{\ell}\right)
{d\over dr}
-\left(M^2+\xi R^{(D+1)} - \frac{m_n^2+\rho^2}{\ell^2\sinh^2(r/\ell)}
\right)\right]u_{n}(r)&=& 0\,.
\label{radeq}
\eea
Here the d'Alembertian $\Box_D$ is for the de Sitter metric,
and we have shifted the 
$D$-dimensional masses to $(m_n^2+\rho^2)H^2$
for later convenience, where 
\beq
\rho^2 H^2=\xi_c R^{(D)}={(D-1)^2H^2\over 4}
\eeq 
with $\xi_c=(D-1)/(4D)$, the conformal curvature coupling constant.
Then, we may quantise each $\phi_n(x)$ as a $D$-dimensional scalar field,
provided that the functions $u_n(r)$ are properly normalised,
as explicitly given below in Eq.~(\ref{radnorm}).
Thus, we have a tower of $D$-dimensional scalar fields instead of
a single $(D+1)$-dimensional scalar field. This is what is conventionally
called Kaluza-Klein or dimensional reduction.
This allows us to express $\langle\phi^2(X)\rangle$ as,
instead of Eq.~(\ref{phi2}),
\begin{eqnarray}
\langle\phi^2(X)\rangle=\sum_n |u_n(r)|^2\langle\phi_n^2(x)\rangle.
\label{drphi2}
\end{eqnarray}
More generally, any kind of Green function\footnote{%
We could also use the Green function to find the vacuum expectation value 
of the stress energy tensor.
} 
(two-point function) can be expressed as
\begin{eqnarray}
G(r,x;r',x')=\sum_n u_n(r)u_n(r')G_n(x,x')\,.
\end{eqnarray}
This is the basic formula we shall use in the present paper.

Now let us explicitly perform the above procedure.
Inserting the decomposition (\ref{drphi}) of $\phi$ into the
 action~(\ref{act}), we find it reduces to
\beq
S=\frac 1 2 \sum_n\int d^{D}x \sqrt{-g^{(D)}}
~\phi_n(x)\left(\Box_{D}-(m_n^2+\rho^2)H^2\right)\phi_n(x),
\eeq
provided that the functions $u_n(r)$ are normalised as
\beq
2\int_{r_-}^{r_+}dr
\,b^{D-2}(r)\,u_{n}(r)u_{n'}(r)= \delta_{nn'}\,.
\label{radnorm}
\eeq
Note that we assume the extra-dimension is compact
with ${\mathbb Z}_2$-symmetry. That is,
there are two identical copies of the bulk at both sides of each brane.
This gives the factor of two in front of the above normalisation condition.

The solution of the bulk radial equation~(\ref{radeq}) is given by
\bea
u_{n}(r) &=&
\frac{1}{N_n}{1\over (H\ell\sinh(r/\ell))^{D/2}}
\left( 
P_{im_n - 1/2}^{-\nu}[\coth(r/\ell)]
+C_n P_{im_n - 1/2}^{\nu}[\coth(r/\ell)]
\right),
\label{sols}
\eea
where $N_n$ is a normalisation constant and
\beq
\nu =\sqrt{{D^2\over 4}+\ell^2M^2-\xi D(D+1) }
\eeq
The solutions given above are equivalent to those obtained by the standard
dimensional reduction procedure, see Ref.~\cite{BLR}.

To consider the quantum fluctuations of $D$-dimensional fields $\phi_n$,
we assume that the vacuum is given by the Euclidean vacuum. In de Sitter
space, this corresponds to choosing the Bunch-Davies vacuum which is
de Sitter invariant. Then the quantum fluctuations are described
by fluctuations in a Euclideanised action, obtained
by an analytic continuation of the metric, see Ref.~\cite{GS},
\beq
ds^2 =dr^2+\ell^2\sinh^2(r/\ell)d\Omega_{D}^2,
\label{Emet}
\eeq
where $d\Omega_{D}^2$ is the metric on the $D$-sphere
with volume
\beq
V_{S^{D}}={2\pi^{{D+1\over 2}}\over\Gamma({D+1\over 2})}\,.
\label{volsD}
\eeq

Furthermore, for calculational purposes, it is convenient to work in 
the conformal frame with the line element defined by, e.g.,
 see Refs.~\cite{GS,GEN},
\bea
ds^2 =a^{2}(z)(dz^2+d\Omega^2_D)\,;
\qquad
a(z)=\ell\sinh(r/\ell)=\frac{\ell}{\sinh(z_0+|z|)}\,.
\label{confmetric}
\eea
In this coordinate system the positive tension brane is located at $z=0$ 
and the other brane with negative tension is placed at $|z|=L$. 
Note that the Hubble constant on each brane is
determined by its location,
\begin{equation}
H_{\pm} =\frac{1}{\ell\sinh(r_{\pm}/\ell)}=
\left\{\begin{array}{l}
\sinh z_0\,,\\
\sinh(z_0+L).
\end{array}\right.
\label{hub}
\end{equation}
The non-dimensional length $L$, 
defining the distance between the two branes, 
is given in terms of the physical length $r$ as
\begin{equation}
L=\int_{r_-}^{r_+}\frac{dr}
{\ell\sinh(r/\ell)}=
\log\coth{r_-\over 2\ell}-\log\coth{r_+\over 2\ell}\,.
\end{equation}

The normalisation~(\ref{radnorm}) in the conformal time frame is then
\beq
\int_{-L}^{L} dz~f_{n}(z)\,f_{n'}(z)= \delta_{nn'}\,,
\label{norm}
\eeq
where we have introduced a function $f_n(z)$ by 
\beq
f_{n}(z)= H^{\rho-1/2}a^{\rho}\,u_{n}(z)\,;\qquad\rho=\frac{D-1}{2}\,.
\eeq
Then
\beq
f_n(z)={1\over \sqrt{a(z)}\,H\, N_n}\left(P^{-\nu}_{im_n-1/2}[\cosh (z_0+|z|)]
+ C_{n}\, P^\nu_{im_n-1/2}[\cosh (z_0+|z|)]\right)\,.
\eeq

It is straightforward to see that the boundary conditions
 in the proper time frame, which are 
\beq
[\partial_r ]^{+}_{-} u_{n}(r) 
=-{2D\xi\over\ell} [\coth(r/\ell)]^+_-u_{n}(r),
\eeq 
lead to the following Robin type boundary conditions in 
the conformal time frame
\begin{equation}
\left[\partial_z+\alpha\right]^{+}_{-}f_n=0\,;
\qquad
\alpha=2D(\xi_c-\xi)\,.
\label{robcs}
\end{equation}
where conformal coupling, $\xi=\xi_c$, implies Neumann boundary conditions. 

In the case of a single positive tension brane, there may or
may not be a bound state mode, depending on the mass and
curvature coupling.
The existence of a bound state mode
is determined from the boundary condition~\cite{HS,LS}.
All the rest of modes are unbounded, giving rise to
a continuous spectrum of Kaluza-Klein modes.
In contrast, in the two brane case,
all the modes are essentially bounded.
Here, however, we call the lowest $n=0$ mode
in the Kaluza-Klein tower the bound state.
In the large separation limit of two branes,
this mode reduces to the bound state mode if
there exists a bound state in the single brane case.

\section{Green Function and Zeta function}
\label{zetphi}

As mentioned already,
with the dimensional reduction performed,
the vacuum expectation value of the 
field fluctuations are given by
\beq
\langle\phi(z,x)\phi(z',x')\rangle
=\sum_n u_n(z) u_n(z')\langle\phi_n(x)\phi_n(x')\rangle\,.
\eeq
In the above, however, the ${\mathbb Z}_2$-symmetry is not properly
taken into account. Since the two points $z$ and $-z$ are identical,
we may focus only on the range of positive $z$; $0\leq z\leq L$.
Then the above two-point function should be expressed as
\begin{eqnarray}
\left.\langle\phi(z,x)\phi(z',x')\rangle\right|_{L\geq z,z'\geq0}
&=&\Bigl(\langle\phi(z,x)\phi(z',x')\rangle
+\langle\phi(z,x)\phi(-z',x')\rangle\Bigr)
\nn
&=&2\sum_n u_n(z) u_n(z')\langle\phi_n(x)\phi_n(x')\rangle\,.
\end{eqnarray}
Similarly, the Green function can also be expressed as a
tower of de Sitter space Green functions (with 
Kaluza-Klein masses $m_n$),
\beq
G(z,x;z',x')=2\sum_n u_n(z) u_n(z')~G_n(x,x')
=\frac{2H}
{H^{2\rho}a^{\rho}(z)a^{\rho}(z')}\sum_n f_n(z) f_n(z')~G_n(x,x')\,,
\label{fgreen}
\eeq
where the factor two is due to the ${\mathbb Z}_2$-symmetry, with the
understanding that we restrict the range of $z$ to be $0\leq z\leq L$.

The de Sitter space Green function can be expressed in 
terms of its associated heat kernel by
\beq
G_n(x,x')=\int_0^\infty dt ~K_n(x,x';t),
\label{sgreen}
\eeq
where the heat kernel is expressed in terms of the
eigenvalues and normalised eigenfunctions for the
field operator in $D$-dimensions,
\beq
\left[-\Box_D+(m_n^2+\rho^2)H^2\right]\phi_{j}(x)=\lambda_j\phi_{j}(x),
\eeq 
as
\begin{eqnarray}
K_n(x,x';t)=\sum_j\phi_j(x)\phi_j(x')e^{-\lambda_jt}\,.
\end{eqnarray}

We briefly mention that for calculations involving global quantities,
 such as the one-loop effective action, the {\it integrated} heat kernel
 is of more interest, which is defined by the functional trace
\beq
 K_n(t)=\int d^Dx ~\sqrt{-g^{(D)}}~ K_n(x,x;t)\,.
\eeq
Thus, for the case of de Sitter space, which is a maximally
symmetric space, the global heat kernel is trivially related to
(the diagonal part of) the local one by a volume
 factor and the diagonal part (i.e., $x=x'$) of the local
heat kernel is given by
\beq
K_n(x,x;t)={H^D\over V_{S^D}}K_n(t)\,;
\qquad K_n(t)=\sum_j d_j ~e^{-\lambda_j t}\,,
\label{skern}
\eeq 
where $d_j$ is the degeneracy of the eigenvalue $\lambda_j$,
and $V_{S^D}$ is the volume of the $D$-sphere, Eq.~(\ref{volsD}).

The d'Alembertian on the $D$-sphere has the spectrum,
\beq
-\Box_D\phi_{j}(x)=j(j+2\rho)H^2\phi_{j}(x)\,,
\eeq 
with associated degeneracy 
\beq
d_j=(2j+D-1){(j+D-2)!\over j!\,(D-1)!}
=\sum_{k=1}^{D-1}~e_k(D) \left(j+\rho\right)^k\,;
\qquad\rho=\frac{D-1}{2}\,.
\label{deg}
\eeq
The coefficients $e_k(D)$ can be found by explicit calculation for 
a given dimension $D$. Thus, the eigenvalues are given by
\beq
\lambda_{j}=\left(m_n^2+\left(j+\rho\right)^2\right)H^2.
\eeq
If the Kaluza-Klein eigenvalues, $m_n$, are explicitly known,
 as in the special case discussed in the next section,
we can perform the mode sums directly.

Of course, the quantities we are interested in need to be regularised and 
we shall use that fact that the local zeta function is related to
 the heat kernel by a Mellin transform, i.e., 
\beq
\zeta_n(s)(x,x')={1\over \Gamma(s)} \int_0^\infty dt~ t^{s-1} K_n(x,x';t)\,,
\label{mell}
\eeq
where the subscript $n$ refers to the zeta function for
 each Kaluza-Klein mode,
with the limit $s\rightarrow 1$ clearly giving the Green function, 
see Eq.~(\ref{sgreen}).
In particular, for $x=x'$, $\zeta_n(s)(x,x)$ will be
independent of $x$ and given by the global zeta function
divided by the volume $V_{S^D}/H^D$:
\begin{eqnarray}
\zeta_n(s)(x,x)=\frac{H^D}{V_{S^D}}\,\zeta_n(s)\,;
\qquad
\zeta_n(s)={1\over \Gamma(s)} \int_0^\infty dt~ t^{s-1} K_n(t)\,.
\end{eqnarray}
Then, we introduce the local zeta function for the Kaluza-Klein tower by
\beq
\zeta(s)(z,z')=\frac{2H^{D+1}}{H^{2\rho}a^{\rho}(z)a^{\rho}(z')V_{S^D}}
 \sum_n ~f_n(z) \,f_n(z')\,\zeta_n(s)\,.
\eeq
Thence, 
\begin{equation}
\langle\phi(z)\phi(z')\rangle =\zeta(s)(z,z')\big|_{s=1}\,.
\label{naive}
\end{equation}
In particular, for $z=z'$, by introducing the diagonal part
of the local heat kernel,
\begin{eqnarray}
K(z,t)\equiv 2\sum_nf_n^2(z)K_n(t)\,,
\label{Kdiag}
\end{eqnarray}
The diagonal part of the local zeta function can be expressed as
\begin{eqnarray}
\zeta(s)(z)&\equiv&
\zeta(s)(z,z)=\frac{2H^{D+1}}{H^{2\rho}a^{2\rho}(z)V_{S^D}}
\sum_nf_n^2(z)\zeta_n(s)
\nonumber\\
&=&\frac{1}{a^{D-1}(z)V_{S^D}}\frac{H^2}{\Gamma(s)}
\int_0^\infty dt\,t^{s-1}K(z,t)\,.
\label{diagonal}
\end{eqnarray}
For some simple cases, such as massless, conformal scalar fields, 
the coincidence limit of the quantum fluctuations are straightforward
enough to evaluate in terms of a local zeta function, 
because the heat kernel is exact. 

\section{A Special Case}
\label{spec}

Our final goal is to quantify the quantum effect of a
bulk scalar field with arbitrary mass and curvature coupling
on the dynamics of the bulk spacetime as well as of the branes.
However, unfortunately, we have no analytical method to deal
with the general case. So, we shall consider a special case that
can be analytically evaluated, hoping that it nevertheless contains
some features that are common to the general case.

Due to the great simplification of the problem,
we shall consider scalar
 fields with $\nu=1/2$, where the solutions reduce to
\beq
f_n(z)=\sqrt{{2\over \pi \ell}}{1\over N_n}
\left({1\over m_n}\sin m_n(z_0+|z|)+ C_{n}\cos m_n(z_0+|z|)\right).
\label{orsol}
\eeq
This solution encompasses not just the conformally invariant case, 
$M=0, ~\xi=\xi_c$, but any choice of mass and curvature coupling obeying
 the relation
\beq
M^2\ell^2-D(D+1)\xi={1-D^2\over 4}\,,
\label{class}
\eeq
which includes minimal coupling, $\xi=0$, for negative mass squared, 
$M^2=(1-D^2)/(4\ell^2)$, which is a valid parameter for the bulk inflaton 
model, see Ref.~\cite{HS}. However, as we will show, see Eq.~(\ref{norsol}), 
only in particular cases for general $\nu=1/2$, e.g., $z_0=L/4$, does the 
solution reduce to a form that allows for a direct mode summation to
 obtain the quantum fluctuations.
 Note, this also changes the value of the bound state contribution.

The boundary conditions~(\ref{robcs}) on each of the branes enables us 
determine the weight of the second linearly independent solution to be
\beq
C_n=-{\cos m_n z_0+\frac \alpha{m_n} \sin m_n z_0\over
\alpha\cos m_n z_0- m_n\sin m_n z_0}=
-{\cos m_n (z_0+L)+\frac \alpha{m_n} \sin m_n (z_0+L)\over
\alpha\cos m_n (z_0+L)-m_n\sin m_n (z_0+L)}\,.
\label{wait}
\eeq
Moreover, the solution of the above implicit eigenvalue equation is
\beq
m_n={n\pi\over L}\,.
\eeq

It is instructive to write Eq.~(\ref{orsol}) in a slightly different form:
\beq
f_n(z)=\sqrt{{2\over \pi \ell }}{1\over m_n N_n}
\left(A_n\cos m_n|z|+B_{n}\sin m_n|z|\right).
\eeq
The normalisation of the radial eigenfunctions
 $f_n(z)$, Eq.~(\ref{norm}), implies
\beq
 N_n=\frac L {\pi\ell} { A_n^2+B_n^2\over m_n^2}\,,
\eeq
and thus the solution can be written as
\beq
f_n(z)=\sqrt{{1\over L}} {1\over \sqrt{A_n^2+B_n^2}}
\big(A_n\cos m_n|z|+B_{n}\sin m_n|z|\big)\,.
\eeq
Then the boundary conditions~(\ref{robcs}) imply
$\alpha\,A_n+m_n B_n=0$. 
Hence,
\beq
f_n(z)=\sqrt{{1\over L}} {1\over \sqrt{m_n^2+\alpha^2}}
\big(m_n\cos m_n|z|-\alpha\sin m_n|z|\big)\,.
\label{norsol}
\eeq
Thus, for Neumann boundary conditions, i.e., $\alpha=0$, 
we obtain the simple normalised solution
\beq
f_n(z)=\sqrt{{1\over L}} ~\cos \left(\frac{n\pi}{L}\,|z|\right)\,;
\qquad n=0,1,2,\cdots.
\label{consol}
\eeq
For our model given by the action~(\ref{act}), 
this corresponds to the case of a massless ($M=0$),
 conformally coupled ($\xi=\xi_c$) scalar. However, it is
worth noting that if we introduce a non-vanishing 
coupling of the scalar field to the tension of the brane, 
the scalar field can be non-conformally coupled, as long as
the relation~(\ref{class}) between $M^2$ and $\xi$ is satisfied.
Another simple case is the case of Dirichlet boundary conditions,
 i.e., $\alpha=\infty$, for which the solution~(\ref{orsol})
 simplifies to the above solution with
 cosine replaced by sine.

As noted earlier, we call the $n=0$ mode the bound state mode in general.
However, for the special case discussed here, let us call the $n=0$
mode the {\it zero mode}, because the eigenfunction $f_0(z)$ is
just a constant.
Note also, that like for the continuous mass spectrum, analysed in 
\cite{SHS}, the discrete Kaluza-Klein spectrum is also 
independent of the higher-dimensional mass of the scalar field, i.e., 
$M^2+\xi R^{(D+1)}$. However, it does depend on the boundary conditions,
 see Eq.~(\ref{wait}), which are determined from the jump in $R^{(D+1)}$
 across each boundary.

\section{Quantum Fluctuations}
\label{flucts}

As we have seen in Eq.~(\ref{diagonal}), 
the quantum fluctuations can be found by working with the local 
heat kernel~(\ref{Kdiag}).
In the case of $\nu=1/2$ with Neumann boundary conditions,
we have
\beq
K(z,t)=2\sum_nf_n^2(z)K_n(t)
={2\over L }
\sum_{n=0}^{\infty}\sum_{j=0}^\infty 
d_j ~e^{-(j+\rho)^2H^2t}e^{-m_n^2H^2t} \cos^2\left(m_n z\right)\,.
\eeq
We comment on the Dirichlet case where appropriate.
 The above heat kernel reduces to the flat brane one when 
the modes $m$, of the $D$-sphere, become continuous.
For Dirichlet boundary conditions the eigenfunctions 
are sine instead of cosine and the mode sum starts from $n=1$.

Employing steps similar to the evaluation 
of the effective action, see Ref.~\cite{NS},
introducing $\tilde\zeta(s)(z)$ by
\begin{eqnarray}
\zeta(s)(z)=\frac{1}{a^{D-1}(z)V_{S^D}}\,H^2\tilde\zeta(s)(z)\,,
\end{eqnarray}
we manipulate as
\bea
\tilde\zeta(s)(z)
&=&{2H^{-2s}\over L\Gamma(s)} \sum_{n=0}^{\infty} \sum_{j=0}^\infty 
d_j \int_0^\infty dt ~t^{s-1} ~e^{-(j+\rho)^2t}e^{-m_n^2t}
 \cos^2\left(m_n z\right)
\\
&=&{2H^{-2s}\over L\Gamma(s)} \sum_{j=0}^\infty 
d_j \int_0^\infty dt ~t^{s-1} ~e^{-(j+\rho)^2t}
\nonumber\\
&&
+{H^{-2s}\over L\Gamma(s)} \sum_{n=1}^{\infty} \sum_{j=0}^\infty 
d_j \int_0^\infty dt ~t^{s-1} ~e^{-(j+\rho)^2t}\, e^{-m_n^2t}\,
\left[1+ \cos\left(2m_n z\right)\right]\,.
\nonumber
\label{zetabeg}
\eea
The above zeta function converges for $s>5/2$ and, to continue to 
the value $s=1$, we can apply the Poisson resummation formula~\cite{WW}
\beq
\sum_{n=1}^{\infty} e^{-\pi^2 n^2 t/L^2}\cos\left({2n\pi z\over L}\right)
=\frac{1}{2}\left(-1+\sqrt{L^2\over \pi t}
\sum_{n=-\infty}^{\infty} e^{-(n L+ z)^2/ t}\right)\,,
\eeq
which implies that 
\bea
\tilde\zeta(s)(z)&=&\frac {\zeta_{\rm D}(s)}{L}+
{H^{-2s}\over 2\sqrt\pi ~\Gamma(s) }
 \sum_{n=-\infty}^{\infty} \sum_{j=0}^\infty 
d_j \int_0^\infty dt ~t^{s-1} ~e^{-(j+\rho)^2t}
\left(e^{-n^2L^2/t}+ e^{-(nL+z)^2/t}\right)\,,
\eea
where
\beq
\zeta_{\rm D}(s)={H^{-2s}\over \Gamma(s) }\sum_{m=0}^\infty 
d_j \int_0^\infty dt ~t^{s-1} ~e^{-(j+\rho)^2t}=H^{-2s}\sum_{j=0}^\infty 
d_j(j+\rho)^{-2s}\,,
\label{zetds}
\eeq
which is nothing but the local zeta function for the $D$-sphere
of radius $H^{-1}$, see Appendix~\ref{bstate}.
 Thus we can regard the first term,
$\tilde\zeta_{\rm zero}\equiv\zeta_{\rm D}/L$,
as the zero mode contribution to the zeta function.
Therefore, by defining the Kaluza-Klein contribution as
$\tilde\zeta_{\rm kk}=\tilde\zeta-\tilde\zeta_{\rm zero}$, we have
\bea
\tilde\zeta_{\rm kk}(s)(z)&=&
{H^{-2s}\over 2\sqrt\pi ~ \Gamma(s)}
 \sum_{n=-\infty}^{\infty}\sum_{j=0}^\infty
d_j \int_0^\infty dt ~t^{s-3/2} e^{-(j+\rho)^2t}
\left[ e^{-n^2 L^2/t} + \frac{1}{2}\left(e^{-(nL+z)^2/t}
+ e^{-(nL-z)^2/t}\right)\right]\nn
\label{manip1}
\eea
where we have written the above in a form that is manifestly symmetric 
under $z\to -z$. Similar considerations for the Dirichlet
 case lead to a minus sign in the front of the last two terms
 in Eq.~(\ref{manip1}). The expression $\tilde\zeta_{\rm kk}$ 
is free of poles and, as shown in Appendix~\ref{bstate}, the only pole 
comes from $\tilde\zeta_{\rm zero}$. 

Let us further proceed to simplify $\tilde\zeta_{\rm kk}$.
 Using the fact that
\beq
\sum_{n=-\infty}^{\infty} e^{-(n \pm a)^2}
=e^{-a^2}+\sum_{n=1}^{\infty} e^{-(n+ a)^2}
+\sum_{n=1}^{\infty} e^{-(n- a)^2}
\eeq
implies 
\bea
\label{surf}
\tilde\zeta_{\rm kk}(s)(z)
&=&{H^{-2s}\over 2\sqrt\pi ~ \Gamma(s)}\sum_{j=0}^\infty
d_j \int_0^\infty dt ~t^{s-3/2} e^{-(j+\rho)^2t}
(1+ e^{-z^2/t})\\
&&+{H^{-2s}\over \sqrt\pi ~ \Gamma(s)}
 \sum_{n=1}^{\infty}\sum_{j=0}^\infty
d_j \int_0^\infty dt ~t^{s-3/2} e^{-(j+\rho)^2t}
\left[ e^{-n^2 L^2/t} + \frac{1}{2}\left(e^{-(nL+z)^2/t}
+ e^{-(nL-z)^2/t}\right)\right].
\nonumber
\eea
It is worth mentioning that if we set $z=0$ in the above equation, which is equivalent 
to constraining the field on the brane from the beginning (see the next subsection), then it is free of surface divergences.
 It is only the process of integration over $t$ in the above expression,
 with the regularising parameter $s$, that brings about the appearance of
 any surface divergences. Thus, at least in this simple case and by way of
 zeta function regularisation, it appears that surface divergences are
 intrinsically related to the regularisation of ultraviolet divergences. 

Integrating over the parameter $t$, and employing the identity 
\beq
\int^\infty_0 dx~ x^{\nu-1}\exp\left[-{\beta\over x} -\gamma x\right]
=2\left({\beta\over\gamma}\right)^{\nu/2}K_\nu (2\sqrt{\beta\gamma})\,,
\eeq
for all but the first term, we find the local zeta function to be:
\bea
H^{2s}\,\tilde\zeta_{\rm kk}(s)(z)
&=&
\frac{\Gamma(s-3/2)}{2\sqrt\pi ~\Gamma(s)}\zeta_{\rm D}(s-1/2)
+\frac{1}{\sqrt\pi~\Gamma(s)}\sum_{j=0}^\infty ~d_j~z^{s-1/2}
(j+\rho)^{-s+1/2}~K_{s-1/2}[2z(j+\rho)]
\nn
&&+\frac{2}{\sqrt\pi~\Gamma(s)}
\sum_{n=1,~j=0}^\infty ~d_j~(nL)^{s-1/2}(j+\rho)^{-s+1/2}~
K_{s-1/2}[2nL(j+\rho)]
\nn
&&+\frac{1}{\sqrt\pi~\Gamma(s)}\sum_{n=1,~j=0}^\infty ~
d_j~(nL+z)^{s-1/2}(j+\rho)^{-s+1/2}~K_{s-1/2}[2(nL+z)(j+\rho)]
\nn
&&+\frac{1}{\sqrt\pi~\Gamma(s)}\sum_{n=1,~j=0}^\infty ~
d_j~(nL-z)^{s-1/2}(j+\rho)^{-s+1/2}~K_{s-1/2}[2(nL-z)(j+\rho)]\,.
\eea
The above expression is a general expression valid for any $s$. 
However, for the purposes of this paper we wish to obtain the vacuum
 expectation value of the square of the field fluctuations, i.e.,
 set $s=1$, implying
\bea
H^{2}\langle\tilde\phi^2(z)\rangle_{\rm kk}
&=&
-\zeta_{\rm D}(1/2)
+\frac{1}{\sqrt\pi}\sum_{j=0}^\infty ~d_j~z^{1/2}(j+\rho)^{-1/2}~
K_{1/2}[2z(j+\rho)]
\nn
&&+\frac{2}{\sqrt\pi }\sum_{n=1,~j=0}^\infty ~d_j~(nL)^{1/2}(j+\rho)^{-1/2}~
K_{1/2}[2nL(j+\rho)]
\nn
&&+\frac{1}{\sqrt\pi }\sum_{n=1,~j=0}^\infty ~d_j~(nL+z)^{1/2}(j+\rho)^{-1/2}~
K_{1/2}[2(nL+z)(j+\rho)]
\nn
&&+\frac{1}{\sqrt\pi }\sum_{n=1,~j=0}^\infty ~d_j~(nL-z)^{1/2}(j+\rho)^{-1/2}~
K_{1/2}[2(nL-z)(j+\rho)]\,,
\eea
where $\langle\tilde\phi^2\rangle$ is defined, 
similarly to $\tilde\zeta(s)$, by
\begin{eqnarray}
\langle\phi^2(z)\rangle=
\frac{1}{a^{D-1}(z)V_{S^D}}\,H^2\langle\tilde\phi^2(z)\rangle\,.
\end{eqnarray}
Using the fact that $K_{1/2}(x)=\sqrt{\pi/(2x)}e^{-x}$, we end up with 
\bea
H^{2}\langle\tilde\phi^2(z)\rangle_{\rm kk}
&=&
-\zeta_{\rm D}(1/2)
+\sum_{n=1,~j=0}^\infty 
{d_j\over  (j+\rho)}~e^{-2nL(j+\rho)}
+\frac 1 2\sum_{j=0}^\infty
{d_j\over  (j+\rho)}~e^{-2z(j+\rho)}
\nn
&&+\frac 1 2\sum_{n=1,~j=0}^\infty 
{d_j\over (j+\rho)}~e^{-2(nL+z)(j+\rho)}
+\frac 1 2\sum_{n=1,~j=0}^\infty 
{d_j\over (j+\rho)}~e^{-2(nL-z)(j+\rho)}\,,
\nn
\label{flatuse}
\eea
where the above expression is for general $(D+1)$-dimensions.
The first two terms are the global parts, i.e., they are constant  
in the $z$-direction, with the first term due to one brane. The other
 three terms are  local functions of $z$, with the third term the
 contribution for just one brane. 

Now we shall specifically look at the case for $D=4$,
 where from Eq.~(\ref{deg}),
\beq
d_j=\frac{1}{3}\left(j+\frac{3}{2}\right)
\left(\left(j+\frac{3}{2}\right)^2+\frac{1}{4}\right)\,.
\eeq
In this case, it is easy to verify that $\zeta_{\rm D}(1/2)=0$ and hence 
the first term  in Eq.~(\ref{flatuse}) is equal to zero. In general,
 $\zeta_{\rm D}(1/2)$ 
appears to be zero for odd dimensions, implying $D$ even, as can be 
easily verified for $D=2,4$.

A very succinct expression can be obtained if we sum over $j$ first, 
\begin{eqnarray}
H^2\langle\tilde\phi^2(z)\rangle_{\rm kk}
=
\frac 1 {12}\sum_{n=1}^\infty 
{1\over \sinh^3(nL)} 
&+&\frac 1 {24}{1 \over   \sinh^3 z}
+\frac 1 {24}\sum_{n=1}^\infty 
{1\over \sinh^3 (nL+z)}
\nonumber\\
&+&\frac 1 {24} {1\over \sinh^3 (L-z)}
+\frac 1 {24}\sum_{n=2}^\infty 
{1\over \sinh^3 (nL-z)}\,,
\label{finres}
\end{eqnarray}
where the second and fourth terms have been separated from the local mode
 sums for reasons which shall shortly become apparent.
Note, for Dirichlet  boundary conditions the signs reverse on all but 
the first term in the above equation. 

Straightaway we are able to obtain the result for when one of the branes
 is absent, $L\rightarrow\infty$,
\beq
H^2\langle\tilde\phi^2(z)\rangle_{\rm kk}
\mathop{\longrightarrow}_{L\to\infty} 
\frac 1 {24}{1 \over \sinh^3 z}\,,
\label{oneres}
\eeq
which is non-zero, unlike the global piece in Eq.~(\ref{flatuse}). 
By symmetry,
 the fourth term gives the quantum fluctuations for a single negative
 tension brane, located at $z=L$. Quite clearly, 
we have run into trouble, because, for example, in the vicinity of the
 positive tension brane, at $z=\varepsilon$ with $0\leq z \leq L$,
 we find surface divergences as we approach the brane boundary,
\beq
H^2\langle\tilde\phi^2(\varepsilon)\rangle_{\rm kk}\sim
\frac 1{24~\varepsilon^3}-\frac 1{48~\varepsilon}+ \,\mbox{finite terms}\,,
\eeq
with a similar result on the negative tension brane at $z=L-\varepsilon$.
Furthermore, these surface divergences are non-integrable, and as is well 
known in the case of the stress energy tensor, they lead to divergences
 in the Casimir Force, even after subtracting ultraviolet effects.

It is interesting to compare this result with the boundary divergences of
 the flat brane limit, which has been evaluated in Appendix~\ref{flatapp}.
 In the vicinity of the brane at $z=\varepsilon$ 
we obtain, from Eq.~(\ref{flatres}),
\beq
H^2\langle\tilde\phi^2(\varepsilon)\rangle^{\rm flat}_{\rm kk}\sim  
{1\over 24~\varepsilon^3} +\,\mbox{finite terms}\,.
\eeq
Thus, to leading order the two results agree, but to next order there
 is a difference due to the curvature of the boundary, e.g.,
 see Ref.~\cite{DC}.

\subsection*{Constrained to the Brane}
\label{const}

Let us now discuss quantum fluctuations on the brane.
Recalling that the bulk inflaton model assumes that
the positive tension brane is our universe, we evaluate
$\langle\phi^2\rangle$ on the positive tension brane.
Putting $z\to0$ in Eq.~(\ref{surf}), we have
\begin{eqnarray}
H^{2s}\,\tilde\zeta_{\rm kk}(s)(0)
&=&{1\over \sqrt\pi ~ \Gamma(s)}\sum_{j=0}^\infty
d_j \int_0^\infty dt ~t^{s-3/2} e^{-(j+\rho)^2t}
\nn
&&+{2\over \sqrt\pi ~ \Gamma(s)}
 \sum_{n=1}^{\infty}\sum_{j=0}^\infty
d_j \int_0^\infty dt ~t^{s-3/2} e^{-(j+\rho)^2t}\,e^{-n^2 L^2/t}.
\nonumber
\end{eqnarray} 
which we wish to evaluate at $s=1$. Then, following steps similar to those
 in Sec.~\ref{flucts}, see Eq.~(\ref{surf}),
 we find the simple result, for $D=4$,
\beq
H^2\langle\tilde\phi^2(0)\rangle^{\rm reg}_{\rm kk}=
\frac {1} {6}\sum_{n=1}^\infty 
{1\over \sinh^3(nL)}\,,
\label{conres}
\eeq
which is free of surface divergences. This is equivalent to
\begin{eqnarray}
\lim_{z\to0}
\left[H^2\langle\tilde\phi^2(z)\rangle_{\rm kk}
-\frac{1}{24}\frac{H^2}{\sinh^3z}
\right],
\end{eqnarray}
i.e., just dropping the surface divergences that appear in the limit 
$z\rightarrow 0$.

\section{Conclusion and Discussion}
\label{conc}

We presented a method to evaluate the vacuum expectation value
$\langle\phi^2(z)\rangle$ in a $(D+1)$-dimensional anti-de Sitter bulk
 bounded by two $D$-dimensional de Sitter branes. 
This background is chosen because of its importance in the scenario
of brane inflation induced by a bulk scalar field, the so-called
bulk inflaton model.
The method is based on a dimensional reduction 
to a lower dimensional theory in de Sitter space satisfying the standard 
canonical commutation relations. We made full use of the fact that
the Euclideanised version of this spacetime is conformal
to the product manifold (or generalised cylinder) $I\times S^D$.
For the evaluation of $\langle\phi^2(z)\rangle$, 
we chose the Euclidean vacuum.
For a restricted class of scalar fields satisfying
 the condition~(\ref{class}), which is essentially
equivalent to a massless, conformal field, we were able to
perform an explicit calculation for $D=4$.
Technically, the evaluation of quantum fluctuations on the
manifold $I\times S^D$ is an interesting calculation in its own right, 
and generalises the calculation of flat to curved boundaries, performed in \cite{IM,CEZ} by zeta function methods. 

For more general cases, where direct mode 
summations are not possible, the evaluation of quantum fluctuations could be
 tackled by Green function approaches, recently used for flat Randall-Sundrum
 branes~\cite{KT,Saha}, for example. Also, see Ref.~\cite{Saha3} for a method more in spirit with the zeta function approach.
The inclusion of other fields, such as a fermion~\cite{MNSS}, would also 
be interesting, particularly with regard to the appearance of surface
 divergences in supersymmetric extensions.
Furthermore, we stress that, like in \cite{KT,Saha,Saha3,SS2}, we are taking into account 
 the local properties of the bulk spacetime, whereas, for example, other works such as \cite{FT,NS,MNSS,SS1,NOZ,ENOO} only considered global quantities for flat or curved branes in de Sitter and anti de Sitter bulks, i.e., the one-loop effective action and related quantities.
 
{}The main result of this paper, for the quantum fluctuations of a massless,
 conformal bulk scalar field is
\beq
\langle\phi^2(z)\rangle_{\rm kk}
=\frac{3\sinh^3(z_0+|z|)}{8\pi^2\ell^3}
H^2\langle\tilde\phi^2(z)\rangle_{\rm kk}\,,
\label{trans}
\eeq
where $H^2\langle\tilde\phi^2(z)\rangle_{\rm kk}$ is
 given by Eq.~(\ref{finres}). This contains all the contributions
of the Kaluza-Klein fields except for the lowest, {\it zero mode}
contribution. The zero mode is equivalent to a 
field with mass-squared $9H^2/4$ on $4$-dimensional de Sitter space
with radius $H^{-1}$, whose result has been well studied~\cite{VF}, but we
recapitulate it in Appendix~\ref{bstate} for completeness. 
Note, that the result given in Eq.~(\ref{trans}) 
can also be applied to the slightly more general
 case of $\nu=1/2$, either for certain values of $z_0$ or for 
certain couplings of the scalar field to the brane tension.

{}From Eq.~(\ref{finres}), it is apparent that even massless, conformal scalar
 fields suffer from surface divergences.\footnote{
As shown for flat branes \cite{KT,Saha}, the stress energy tensor is expected to be free of surface divergences for massless, conformal fields.}
 However, it is commonly accepted
 that these divergences are not a serious problem.
 For example, it can be argued
 that at high energy scales there may be a physical cutoff to
 surface divergences~\cite{DC}, or from a different viewpoint that it is
 possible to renormalise  surface (on-boundary) terms in the effective
 action~\cite{KCD}, see also the comments in Ref.~\cite{Saha2}.
 Rather than this, we could assume that the brane is of a finite thickness,
 where the width of the brane acts as a natural cut-off, but this is
 a significant departure from the original model (another alternative is to apply 
the approach advocated in \cite{FSv}). Certainly, more investigation concerning the issue of surface divergences, in the brane context, is required.

In contrast to the result in the bulk, if we evaluate $\langle\phi^2\rangle$
by constraining the field point on the brane from the beginning,
we find the result is finite. For example, if we take the
field point on the positive tension
 brane (see Sec.~\ref{const}), the result is equivalent to subtracting 
the term responsible for surface divergences from $\langle\phi^2\rangle$ 
in the bulk. The final result on the positive tension brane is
\beq
\langle\phi^2(0)\rangle^{\rm reg}_{\rm kk}
=\frac {\sinh^3z_0}{6\ell^3V_{S^4}}
\sum_{n=1}^\infty \frac{1}{\sinh^3(nL)}
=\frac {H_{+}^3}{16\pi^2}\sum_{n=1}^\infty {1\over \sinh^3(nL)}\,,
\eeq
where $H_{+}$ is the Hubble parameter of the positive tension brane.
Clearly the Kaluza-Klein mode contribution diminishes in the
 limit of large $L$. 
 
We would like to know the relative magnitude of the Kaluza-Klein
 contribution as compared to the zero mode.
The expectation value $\langle\phi^2\rangle$ for the zero mode 
is given in Eq.~(\ref{s4final}).
We find that the ratio of the Kaluza-Klein contribution to the 
zero mode contribution on the positive tension brane is
given by
\beq
\frac{\langle\phi^2(0)\rangle_{\rm kk}}{\langle\phi^2(0)\rangle_{\rm zero}}
=\left(\frac{1}{2}\psi(3/2)
-\frac{11}{12}+\frac{1}{4}\ln[H^2/\mu^2]\right)^{-1}
\sum_{n=1}^\infty {L\over \sinh^3(nL)}\,.
\label{ratio}
\eeq
 Apparently, for large $L$, the Kaluza-Klein contribution becomes
exponentially small. Thus, for a massless, conformal scalar field, 
we find that the Kaluza-Klein contribution is essentially zero
for large $L$, regardless of whether $H_{+}\ell$ is large or small.

From the point of view of the positive tension brane, we may regard the limit
$L\to\infty$ as the one-brane limit. Then, our result shows that
$\langle\phi^2\rangle\to0$ for $L\to\infty$, indicating that
the effect of quantum corrections will be minimal 
(if not totally negligible) for a massless,
conformal bulk scalar field. 

Here, some caution concerning the one-brane limit is in order,
which is quite particular to a de Sitter brane embedded in an 
anti-de Sitter background. When Euclideanised, this space is compact.
A conventional method to evaluate 
global {\it integrated} quantities, such as the one-loop effective
action, is to employ a conformal transformation of the metric to
make the field operator easy to deal with. Then, one evaluates the
cocycle function to compensate the change of path integral measure
due to the conformal transformation. In the two-brane case,
this is a perfectly legitimate procedure~\cite{GaPuTa,GaPuTa2,MNSS,Norm}.
 However, as has been highlighted recently in \cite{FKN},
 in the case of a single de Sitter brane the conformal transformation
alters the topology of the space. As seen from the metric
in conformal coordinates, Eq.~(\ref{confmetric}), 
the conformal coordinate $z$ extends to infinity in the one-brane
case, making the conformally transformed space non-compact.
One would not then expect the effective action on each background to
agree (even after including the cocycle function), as has been explicitly
demonstrated for massless, conformal fields, \cite{FKN}. 

A similar problem arises for our case of $\langle\phi^2\rangle$ as well,
in which we have employed the dimensional reduction method.
It essentially makes use of the same
conformal transformation to define the Kaluza-Klein fields. 
Now, if we try to discuss the one-brane case from the beginning,
we find the Kaluza-Klein spectrum becomes continuous, which 
makes it impossible to define the lower dimensional
Kaluza-Klein fields. In contrast, if we start with the two-brane
case, and take the limit $L\to\infty$ in the end, 
we will be able to obtain a certain result.
It is however totally unclear if this result is anything to do
with the actual one-brane case. To resolve this issue 
we need to develop a different, new technique to deal with
the one-brane case properly.

\section*{Acknowledgments}
We have benefited from stimulating discussions with A. Flachi, Y.~Himemoto,
 A. Knapman, O. Pujol{\`a}s and T.~Tanaka. W.N. is supported under
 the fellowship program, Grant-in-Aid for the 21st Century COE, ``Centre for Diversity and Universality in Physics,'' {\bf @} Kyoto University. M.S. is supported
 in part by Monbukagakusho Grant-in-Aid for Scientific Research(s)
 No. 14102004. 
 
\appendix
\section{Zero Mode}
\label{bstate}

We now wish to evaluate the zero mode contribution on the positive 
tension de Sitter brane, which is just given by
the zeta function for the $D$-sphere. Note that we call this 
$n=0$ mode the zero mode from the conformal frame point of view.
{}From the physical frame point of view, it is a massive mode
with mass given by $\rho^2H^2=(D-1)^2H^2/4$.
  The zero mode contribution
 can be evaluated by other means, e.g., see Ref.~\cite{VF}. 
Here we take the zeta function approach to be consistent with
the main text. 

If we use the zeta function approach a complication arises
 with the zero mode because of the presence of a pole at $s=1$.
 A way around this problem has been discussed in \cite{IM,CEZ}, where
 by employing functional methods, they obtain a more general form for the
 local zeta function 
\bea
\label{modi}
\langle\phi^2(x)\rangle
&=&\mu^{-2}\lim_{s\to1}
\frac{d}{ds}\left[\mu^{2s}(s-1)\,\zeta(s)(x)\right]
\,,
\eea
which agrees with the usual definition when there is no pole at $s=1$.

Focusing on the $5$-dimensional case for a massless conformal scalar field, 
from Eq.~(\ref{zetds}), the zero mode contribution is found to be
\beq
\zeta_{\rm D}(s)=\frac{1}{3 H^{2s}}\left[\zeta_H(2s-3,3/2)
-\frac{1}{4}\,\zeta_H(2s-1,3/2)\right]\,,
\label{zetaD}
\eeq
where $\zeta_H(x,a)$ is the Hurwitz zeta function,
\begin{eqnarray}
\zeta_H(z,a)=\sum_{n=0}^\infty\frac{1}{(n+a)^z}\,.
\end{eqnarray}
It is clear that the second term of Eq.~(\ref{zetaD})
has a pole at $s=1$, given the
following behaviour of the Hurwitz zeta function
\beq
\zeta_H\big(2s-1,3/2\big)=
\frac{1}{2(s-1)}-\psi(3/2)+{\cal O}\big(2(s-1)\big)\,,
\eeq
where
\beq
\psi(3/2)=-\gamma+2-2\ln 2=0.03649....\,,
\eeq
and $\gamma=0.57721....$ is Euler's constant.
However, by way of the {\it improved} zeta function, 
defined above in Eq.~(\ref{modi}), 
we obtain
\begin{eqnarray}
H^2\langle\tilde\phi^2\rangle_{\rm zero} &=&
\frac{1}{L}\,\frac{H^2}{\mu^2}\lim_{s\to1}\frac{d}{ds}
\left[\mu^{2s}(s-1)\,\zeta_D(s)\right]
\nn
&=&\frac{1}{L}\left(\frac{\zeta_H(-1,3/2)}{3}
-\frac{1}{12}\lim_{s\to1}\frac{d}{ds}
\left[\left(\frac{\mu^2}{H^2}\right)^{s-1}
(s-1)\,\zeta_H(2s-1,3/2)\right]\right)
\nn
&=&\frac{1}{L}\left(-\frac{11}{72}+\frac{1}{12}\,\psi(3/2) 
+\frac{1}{24}\,\ln[H^2/\mu^2]\right)\,.
\label{s4half}
\end{eqnarray}
If we multiply the above by $H^2/V_{S^4}=3H^2/(8\pi^2)$,
it agrees with the result for the Bunch-Davies vacuum given in \cite{VF},
for a minimally coupled scalar with 4-dimensional mass-squared $9H^2/4$.

The final result in terms of the physical amplitude of the
scalar field is
\begin{eqnarray}
\langle\phi^2\rangle_{\rm zero}
=\frac{1}{L}\frac{\sinh^3(z_0+|z|)}{\ell^3}\frac{1}{32\pi^2}
\left(-\frac{11}{6}+\psi(3/2) 
+\frac{1}{2}\,\ln[H^2/\mu^2]\right)\,.
\label{s4final}
\end{eqnarray}
It may be noted that the zero mode contribution is inversely
proportional to $L$, while the Kaluza-Klein contribution vanishes
exponentially for $L\to\infty$.

\section{Flat brane limit}
\label{flatapp}

A useful check of our method is to verify that the the flat brane limit
 agrees with well known results, calculated in \cite{IM,CEZ}
 by zeta function methods. In this section all results are
 given in $(D+1)$-dimensions. 

Our starting point is Eq.~(\ref{flatuse}), which for Neumann boundary
 conditions is given by
\bea
H^2\langle\tilde\phi^2(z)\rangle_{\rm kk}&=&
\sum_{n=1,~j=0}^\infty 
{d_j\over\left(j+\rho\right)}~e^{-2nL(j+\rho)}
+\frac 1 2\sum_{j=0}^\infty 
{d_j\over\left(j+\rho\right)}~e^{-2z(j+\rho)}\nn
&&
+\frac 1 2\sum_{n=1,~j=0}^\infty 
{d_j\over \left(j+\rho\right)}~e^{-2(nL+z)(j+\rho)}
+\frac 1 2\sum_{n=1,~j=0}^\infty 
{d_j\over \left(j+\rho\right)}~e^{-2(nL-z)(j+\rho)}\,.
\eea
Summing over $n$ leads to
\bea
H^2\langle\tilde\phi^2(z)\rangle_{\rm kk}&=&
\sum_{j=0}^\infty 
{d_j\over\left(j+\rho\right)}{e^{-2L(j+\rho)}\over1-e^{-2L(j+\rho)}}
+\frac 1 2\sum_{j=0}^\infty {d_j\over  (j+\rho)}~e^{-2z(j+\rho)}\nn
&&
+\frac 1 2\sum_{j=0}^\infty {d_j\over\left(j+\rho\right)}
{e^{-2(L+z)(j+\rho)}\over 1-e^{-2L(j+\rho)}}
+\frac 1 2\sum_{j=0}^\infty 
 {d_j\over\left(j+\rho\right)} 
{ e^{-2 (L-z)(j+\rho)} \over 1- e^{-2L(j+\rho)}}\,.
\eea
The flat brane limit corresponds to $L\ll 1$, because the eigenvalues
$j$ are for the $D$-sphere of unit radius.
In this limit, the modes $m$ are continuous implying the above sums
 become integrals, and therefore, using Eq.~(\ref{deg}),
\beq
{d_j\over\left(j+\rho\right)}\approx{2 j^{D-2}\over (D-1)!}\,,
\eeq
we obtain
\bea
H^2\langle\tilde\phi^2(z)\rangle^{\rm flat}_{\rm kk}&=&
{2 \over (D-1)!}\int_{0}^\infty dj {j^{D-2}e^{-2Lj}\over 1- e^{-2Lj}}
+{1\over (D-1)!}\int_{0}^\infty dj~{j^{D-2}}~e^{-2zj}
\nn
&&+{1\over (D-1)!}\int_{0}^\infty dj{j^{D-2} ~e^{-2(L+z)j}\over 1- e^{-2Lj}}
+{1\over (D-1)!}\int_{0}^\infty dj{j^{D-2} ~e^{-2(L-z)j}\over 1- e^{-2Lj}}\,.
\eea
Simple manipulations then lead to, substituting $x=2Lj$,
\bea
H^2\langle\tilde\phi^2(z)\rangle^{\rm flat}_{\rm kk}&=&
{2 \over (D-1)!} \frac 1{(2L)^{D-1}}\int_{0}^\infty {x^{D-2}\over e^{x}-1}~dx
+{1\over (D-1)!}{(D-2)!\over (2z)^{D-1}}
\\
&&+{1\over (D-1)!} \frac 1{(2L)^{D-1}}\int_{0}^\infty
{x^{D-2} ~e^{-(1+z/L)\,x}\over 1- e^{-x}} dx
+{1\over (D-1)!} \frac 1{(2L)^{D-1}}\int_{0}^\infty
{x^{D-2} ~e^{-(1-z/L)\,x}\over 1- e^{-x}} dx\nonumber\,.
\eea
The remaining integrals can be expressed in terms of Riemann
 and Hurwitz zeta functions~\cite{Grad}:
\bea
H^2\langle\tilde\phi^2(z)\rangle^{\rm flat}_{\rm kk}
&=&
{2 \over (D-1)!} {(D-2)!\over (2L)^{D-1}}~\zeta_R(D-1)
+{1\over (D-1)!}{(D-2)!\over (2z)^{D-1}}
\nn
&&+{1\over (D-1)!} {(D-2)!\over (2L)^{D-1}}~\zeta_H(D-1,1+z/L)
+{1\over (D-1)!} {(D-2)!\over (2L)^{D-1}}~\zeta_H(D-1,1-z/L)\,,
\eea
which can be written in a form suitable for comparison with \cite{CEZ},
\bea
H^2\langle\tilde\phi^2(z)\rangle^{\rm flat}_{\rm kk}&=&
{\Gamma\left({D-1\over 2}\right)\over  (4\pi)^{D-1\over 2}L^{D-1}}
\Big( 2\zeta_R(D-1)+\zeta_H(D-1,z/L)+ \zeta_H(D-1,1-z/L)\Big)\,,
\label{flatres}
\eea
where we divided by the volume of the unit $D$-sphere, $V_{S^{D}}$,
 see Eq.~(\ref{volsD}).
For Dirichlet boundary conditions the signs of all but the first term 
in Eq.~(\ref{flatres}) reverse.
 Thus, we find perfect agreement with the result
 obtained in \cite{CEZ}, quoted for Dirichlet boundary conditions
 (substituting $D=d-1$), which also agrees with the case of $D=3$ given
 in \cite{IM}. For $D=1$ 
there is a pole and we must use a more general definition of the zeta
 function, discussed in Appendix~\ref{bstate}, also see Ref.~\cite{IM,CEZ}.



\end{document}